\documentclass[aps,pra,notitlepage,twocolumn,showpacs,superscriptaddress]{revtex4-1}
\usepackage{amsfonts}
\usepackage{}
\usepackage{amssymb}  % for review and submission
\usepackage{blindtext}
\usepackage{graphicx}  % needed for figures
\usepackage{dcolumn}   % needed for some tables
\usepackage{amsmath,bm,amssymb,latexsym,galois,amsthm}   % for math
\usepackage{dsfont} %数学字体
\usepackage{mathrsfs}
\usepackage[all,cmtip]{xy}% 交换图
\usepackage{bm}
\theoremstyle{plain}% 命令，通过从三个预定义样式中选择其一来定义定理的外观，三个样式分别为：definition（标题粗体，内容罗马体），plain（标题粗体，内容斜体）和remark （标题斜体，内容罗马体）。
\theoremstyle{definition}

\newtheorem*{defn*}{Definition}

\theoremstyle{remark}

\begin{document}

% The following information is for internal review, please remove them for submission
\widetext
%\leftline{Classification: Physics}
%\leftline{To be submitted to (PRL, PRD-RC, PRD, PLB; choose one.)}
%\leftline{Comment to {\tt d0-run2eb-nnn@fnal.gov} by xxx, yyy}
%\mathcal{\mathcal{}}\centerline{\em D\O\ INTERNAL DOCUMENT -- NOT FOR PUBLIC DISTRIBUTION}

% the following line is for submission, including submission to the arXiv!!
%\hspace{5.2in} \mbox{Fermilab-Pub-04/xxx-E}

\title{Geometric Steering Criterion for Two-qubit States}
%\input author_list.tex       % D0 authors (remove the first 3 lines
                             % of this file prior to submission, they
                             % contain a time stamp for the authorlist)
                             % (includes institutions and visitors)
\author{Bai-Chu Yu}
%\email{Email address: yubaichu@mail.ustc.edu.cn}
\affiliation{Key Laboratory of Quantum Information, Chinese Academy of Sciences, School of Physics, University of Science and Technology of China, Hefei, Anhui, 230026, P. R. China}
\affiliation{Synergetic Innovation Center of Quantum Information and Quantum Physics, University of Science and Technology of China, Hefei, Anhui, 230026, P. R. China}

\author{Zhih-Ahn Jia}
%\email{Email address: cajia@mail.ustc.edu.cn}
\affiliation{Key Laboratory of Quantum Information, Chinese Academy of Sciences, School of Physics, University of Science and Technology of China, Hefei, Anhui, 230026, P. R. China}
\affiliation{Synergetic Innovation Center of Quantum Information and Quantum Physics, University of Science and Technology of China, Hefei, Anhui, 230026, P. R. China}

\author{Yu-Chun Wu}
\email{Email address: wuyuchun@ustc.edu.cn}
\affiliation{Key Laboratory of Quantum Information, Chinese Academy of Sciences, School of Physics, University of Science and Technology of China, Hefei, Anhui, 230026, P. R. China}
\affiliation{Synergetic Innovation Center of Quantum Information and Quantum Physics, University of Science and Technology of China, Hefei, Anhui, 230026, P. R. China}

\author{Guang-Can Guo}\affiliation{Key Laboratory of Quantum Information, Chinese Academy of Sciences, School of Physics, University of Science and Technology of China, Hefei, Anhui, 230026, P. R. China}
\affiliation{Synergetic Innovation Center of Quantum Information and Quantum Physics, University of Science and Technology of China, Hefei, Anhui, 230026, P. R. China}

\date{\today}

\begin{abstract}
According to the geometric characterization of measurement assemblages and local hidden state (LHS) models, we propose a steering criterion which is both necessary and sufficient for two-qubit states under arbitrary measurement sets. A quantity is introduced to describe the required local resources to reconstruct a measurement assemblage for two-qubit states. We show that the quantity can be regarded as a quantification of steerability and be used to find out optimal LHS models. Finally we propose a method to generate unsteerable states, and construct some two-qubit states which are entangled but unsteerable under all projective measurements.
\end{abstract}

\maketitle

%\tableofcontents
%\section{\label{sec:level1}First-level heading}
\section{introduction}
\indent The concept of quantum steering was initially introduced by Schr\"{o}dinger \cite{b1}. Consider two distant parties, Alice and Bob, sharing a pair of particles, by performing measurements on her own particle, Alice can steer the particle of Bob into different quantum states. It was recently recognized as a special type of quantum nonlocality intermediate between entanglement and Bell nonlocality \cite{b2,b3}. Different from entanglement and Bell nonlocality, quantum steering is asymmetric between the two parties \cite{b4}. The party which to be steered (say Bob) examines the conditioned states of his own system while receiving the outcomes of the other party's measurement. To make sure that his system is genuinely influenced by Alice's measurements instead of some preexisting local hidden states (LHS), Bob must exclude the LHS model:
\begin{equation}
p(a,b|A,B;W)=\int p(a|A,\lambda)p(b|B,\rho_{\lambda})q(\lambda)d\lambda,
\end{equation}
in which $p(a,b|A,B;W)=\mathrm{Tr}(A_{a}\otimes B_{b} W)$ is the probability of getting outcomes $a$ and $b$ when measurements $A$ and $B$ are performed on $W$ by Alice and Bob respectively, $A_{a}$ and $B_{b}$ are their corresponding measurement operators. $\lambda$ is the local hidden variable, $q(\lambda)$ is the probability density of $\lambda$, and $p(a|A,\lambda)$ is the conditioned probability of Alice obtaining outcome $a$ under $\lambda$. $p(b|B,\rho_{\lambda})=\mathrm{Tr}(B_{b}\rho_{\lambda})$ is the conditioned probability that Bob obtains outcome $b$ while performing $B$ on $\rho(\lambda)$, $\rho(\lambda)$ are some local quantum states on Bob's side determined by $\lambda$.\\
\indent The probability model of LHS in Eq.~(1) can also be expressed in the form:
\begin{equation}
\tilde{\rho}_{A}^{a} =\int p(a|A,\lambda)\rho_{\lambda}q(\lambda)d\lambda,
\end{equation}
where $\tilde{\rho}_{A}^{a}=\mathrm{Tr}_{A}(A_{a}\otimes \bm{I}_{B}W)$, which is the \emph{unnormalized} conditioned state on Bob's side after Alice obtaining $a$ for measurement $A$, the set $\{\tilde{\rho}_{A}^{a}\}$ for all $a$ and $A$ is referred to as a measurement assemblage \cite{b6}. The probability distributions in Eq. (2) should satisfy
\begin{align}
\sum_{a}p(a|A,\lambda)=1\nonumber, &\int q(\lambda)d\lambda=1,\\
\int p(a|A,\lambda)q(\lambda)d&\lambda=p(a|A),
\end{align}
where $p(a|A)=\mathrm{Tr}(\tilde{\rho}_{A}^{a})$ is the probability of getting outcome $a$ when Alice performs measurement $A$.\\
\indent A bipartite state is steerable from Alice to Bob if and only if there is no LHS model $\{p(a|A,\lambda),q(\lambda),\rho_{\lambda}\}$ such that both Eqs. (2) (3) hold for all results $a$ of measurements $A\in \mathfrak{M}_{A}$, where $\mathfrak{M}_{A}$ is the set of measurements Alice is able to perform on her side \cite{b2}.\\
\indent To judge if a state is steerable under certain sets of measurement, we need to check if there is an LHS model for the corresponding measurement assemblage. Many criteria have been proposed, such as inequalities based on the convexity of steering assemblages \cite{b5} or using semidefinite programming (SDP) \cite{b6,b7}, which can be used for a wide range of bipartite states. But still there is no calculable method that can determine the steerability for a generic bipartite state under arbitrary measurement sets even in two-qubit case yet.\\
 \indent Inspired by several works on geometric illustration of quantum steering \cite{b8,b9}, we propose a geometric characterization of measurement assemblages and LHS models, and define a quantity that describes the required local resources to realize the measurement assemblages. We use the characterization and the quantity to (\textrm{i}) obtain a criterion of steering which is both sufficient and necessary under any measurement sets, (\textrm{ii}) define the optimal LHS model for an assemblage, (\textrm{iii}) quantify steerability, (\textrm{iv}) provide a method to generate unsteerable states. Then some examples and further discussions are given.\\
\indent In Ref. \cite{b8}, a geometric representation was given for the measurement assemblage of a two-qubit state under all POVMs. Under the Pauli basis $\sigma_{i}=\{\bm{I},\sigma_{x},\sigma_{y},\sigma_{z}\}$, $i=0,1,2,3$, every single-qubit state $\rho$ can be written as $\rho=\frac{1}{2}(\bm{I}+\bm{b}\cdot\bm{\sigma})$, where the real Bloch vector $\bm{b}=(b_{1},b_{2},b_{3})$ satisfying $\sum_{i=1}^{3}b_{i}^{2}\leq 1$ and $\bm{\sigma}=(\sigma_{1},\sigma_{2},\sigma_{3})$. If we place the Bloch vectors $\bm{b}_{A}^{a}$ of the conditioned states ${\rho}_{A}^{a}=\frac{1}{2}(\bm{I}+\bm{b}_{A}^{a}\cdot\bm{\sigma})$ into a Bloch sphere, a steering ellipsoid is obtained. This method intuitively characterizes the conditioned states under all possible measurements. However, unnormalized states $\tilde{\rho}_{A}^{a}$ have different normalizing factor $p(a|A)$. If we normalize $\tilde{\rho}_{A}^{a}$ in Eq. (2), the corresponding local states $\rho_{\lambda}$ would become unnormalized $\tilde{\rho_{\lambda}}$ with trace $\frac{1}{p(a|A)}$, and a $\tilde{\rho_{\lambda}}$ shared by different $\rho_{A}^{a}$ may have different trace simultaneously, which makes it difficult to geometrically characterize the relation between measurement assemblages and local states.
\section{The geometric model}
\indent In our work, in order to characterize the relation in Eq. (2) efficiently, we put the shrinked Bloch vectors $\bm{s}_{A}^{a}$ into a unit sphere instead of $\bm{b}_{A}^{a}$. $\bm{s}_{A}^{a}$ are vectors satisfying $\bm{s}_{A}^{a}=p(a|A)\cdot\bm{b}_{A}^{a}$, and they correspond to $\tilde{\rho}_{A}^{a}$ as
\begin{equation}
\tilde{\rho}_{A}^{a}=p(a|A){\rho}_{A}^{a}=\frac{1}{2}[p(a|A)\bm{I}+\bm{s}_{A}^{a}\cdot\bm{\sigma}].
\end{equation}
The unit sphere we put $\bm{s}_{A}^{a}$ in is not a Bloch sphere, since the vectors in Bloch sphere represent quantum states with trace 1, while $\bm{s}_{A}^{a}$ represent unnormalized states with trace $p(a|A)$. The length of the vector in this sphere is the product of the trace and the purity of the \emph{unnormalized} conditioned state. Similar to Bloch sphere, the surface of this sphere is comprised of pure qubit states. We would call such sphere a probability Bloch sphere and denote it with $\tilde{B}$, its surface with $N$ and its center with $C$. \\
\indent A two-qubit state $\rho$ can be written in Pauli bases as $\rho= \frac{1} {4}\sum_{u,v=0}^{3} G_{uv} \sigma_{u}\otimes\sigma_{v}$, where $G_{uv}$ is the element of real matrix
$G=\begin{pmatrix}  1&\bm{b}^{t}\\ \bm{a}&\bm{T} \end{pmatrix}$ , $\bm{a}$ and $\bm{b}$ are Bloch vectors, $\bm{T}$ is a $3\times 3$ matrix, and superscript $t$ means transposition. When Alice's side is projected onto a pure state $A_{a}=\frac{1}{2}(\bm{I}+\bm{x}_{A}^{a}\cdot\bm{\sigma})$, Bob's conditioned state becomes
\begin{equation}
\tilde{\rho}_{A}^{a}=\frac{1}{4}[(1+\bm{x}_{A}^{a}\cdot\bm{a})\bm{I}+(\bm{b}+\bm{T}^{t}\bm{x}_{A}^{a})\cdot\bm{\sigma}].
\end{equation}
\indent By comparing equations (4) and (5) we have $p(a|A)=\frac{1}{2}(1+\bm{x}_{A}^{a}\cdot\bm{a})$ and $\bm{s}_{A}^{a}=\frac{1}{2}(\bm{b}+\bm{T}^{t}\bm{x}_{A}^{a})$. We can see that the geometric figure Bob obtains in $\tilde{B}$ under projective measurements of Alice's side is shaped by $\frac{1}{2}\bm{T}^{t}\bm{x}_{A}^{a}$, translated by $\frac{1}{2}\bm{b}$ and independent of $\bm{a}$. The steering figures could be ellipsoids, ellipses, segments and points, the surfaces of them are generated by projective measurements of Alice's side and the inner parts are by POVMs, since any POVM operator can be written as a mixture of some projectors. \\
\indent Hereinafter we call a set $\{p(a|A,\lambda),q(\lambda),\rho(\lambda)\}$ satisfying Eqs. (2) and (3) an LHS model for $\rho$. Now we are ready to construct a geometric characterization of LHS models under arbitrary measurement sets. We will demonstrate the case of one way steering from Alice to Bob, the case for opposite direction is similar.\\
\indent Consider an ensemble $\{q(\lambda),\rho(\lambda)\}$, where $\rho(\lambda)=\frac{1}{2}(\bm{I}+\bm{b}_{\lambda}\cdot \bm{\sigma})$, $\bm{b}_{\lambda}$ is the Bloch vector of $\rho(\lambda)$ which is determined by $\lambda$. Let $\bm{\eta}$ denote a vector in $\tilde{B}$ and $q(\bm{\eta})$ denote a nonnegative distribution of $\bm{\eta}$. To characterize this ensemble in sphere $\tilde{B}$, we let $q(\bm{\eta})=\sum_{\{\lambda|\bm{b}_{\lambda}=\bm{\eta}\}}q(\lambda)$. The sum on the right is over all $\lambda$ which have the same Bloch vectors $\bm{b}_{\lambda}=\bm{\eta}$ (same local state with different distributions). By this method we characterize the states $\rho_{\lambda}$ with vectors $\bm{\eta}$ which satisfy $\bm{\eta}=\bm{b}_{\lambda}$, then we can construct a geometric model that characterizes the LHS model $\{p(a|A,\lambda),q(\lambda),\rho(\lambda)\}$, which we abbreviate as a g-model.\\
\indent$\bf{Definition\ 1.}$  A g-model $\mathcal{G}$ for a steering figure is \emph{nonnegative} distributions $\{q(\bm{\eta})$, $p(a|A,\bm{\eta})\}$ satisfying \\
\noindent(1) Equations
\begin{align}
&\sum_{a}p(a|A,\bm{\eta})=1,\nonumber\\
\int_{\tilde{B}} q(\bm{\eta})p(a|&A,\bm{\eta})d\bm{\eta}=p(a|A)\cdot\int_{\tilde{B}} q(\bm{\eta})d\bm{\eta}
\end{align}
hold for all $a$ and $A$, where the integral is over sphere $\tilde{B}$, $\bm{\eta}$ are vectors in $\tilde{B}$ distributed with $q(\bm{\eta})$. Note that in this work we use $q$ to denote the probability density and $p$ to denote the probability.\\
(2) Equation
\begin{equation}
\bm{s}_{A}^{a}= \int _{\tilde{B}} p(a|A,\bm{\eta})q(\bm{\eta})\bm{\eta}d\bm{\eta}
\end{equation}
holds for all $a$ and $A$.\\
\indent Now we introduce a quantity $\mathbb{S}$ for every g-model.\\
\indent$\bf{Definition\ 2.}$ Let $\mathbb{S}$ denote the integral $\int_{\tilde{B}}q(\bm{\eta})d\bm{\eta}$ for a g-model $\mathcal{G}$.\\
\indent Note that there may be more than one g-models for a steering figure of a two-qubit state, we let the models with the smallest quantity $\mathbb{S}$ among them be the optimal g-models of the figure.\\
\indent$\bf{Definition\ 3.}$ An optimal g-model $\mathcal{G}_{o}$ for a steering figure is the g-model with $\mathbb{S}_{o}$=min$_{i}\{\mathbb{S}_{i}\}$, where $\mathbb{S}_{i}$ is the quantity $\mathbb{S}$ of some g-model $\mathcal{G}_{i}$ for the steering figure.\\
\indent Before making further explanations for the g-model, we propose a steering criterion using g-model.\\
\indent$\bf{Criterion\ 1.}$ An measurement assemblage of two-qubit quantum state admits an LHS model if and only if $\mathbb{S}_{o}\leq1$ for its steering figure.\\
\noindent \emph{Proof}.
For the necessity, the LHS model of the assemblage can be written as $\{p(a|A,\lambda),q(\lambda),\rho(\lambda)\}$, where $\rho(\lambda)=\frac{1}{2}(\bm{I}+\bm{b}_{\lambda}\cdot \bm{\sigma})$ in Pauli basis, $\bm{b}_{\lambda}$ is the Bloch vector determined by $\lambda$, let
\begin{align}
q(\bm{\eta})&=\sum_{\{\lambda|\bm{b}_{\lambda}=\bm{\eta}\}}q(\lambda)\nonumber,\\
p(a|A,\bm{\eta})=&\frac{\sum_{\{\lambda|\bm{b}_{\lambda}=\bm{\eta}\}}p(a|A,\lambda)q(\lambda)}{q(\bm{\eta})}.
\end{align}
then Eqs. (6) and (7) would hold for $\bm{s}_{A}^{a}$ of all $\tilde{\rho}_{A}^{a}$ in Eq. (2). Therefore $\{q(\bm{\eta})$, $p(a|A,\bm{\eta})\}$ is a g-model $\mathcal{G}$ for the steering figure. Now that $\mathbb{S}=\int_{\tilde{B}}q(\bm{\eta})d\bm{\eta}=\int q(\lambda)d\lambda=1$, $\mathbb{S}_{o}\leq \mathbb{S}=1$.\\
\indent For the sufficiency, let $p_{o}(a|A,\bm{\eta})$ and $q_{o}(\bm{\eta})$ denote the distributions of the optimal g-model for the steering figure. When $\mathbb{S}_{o}\leq 1$, we can construct the modified distributions $\{q_{o}'(\bm{\eta}),p_{o}'(a|A,\bm{\eta})\}$ satisfying
\begin{align}
&q_{o}'(\bm{\eta})=(1-\mathbb{S}_{o})\cdot\delta(\bm{\eta})+q_{o}(\bm{\eta}),\nonumber\\
p_{o}'(a|A,&\bm{\eta})=\left \{
\begin{aligned}
&\frac{p(a|A)-I(a|A)}{\int_{C}q_{o}'(\bm{\eta})d\bm{\eta}}, &\quad \bm{\eta}=\bm{0},\\
&p_{o}(a|A,\bm{\eta}), &\quad otherwise,
\end{aligned}
\right.
\end{align}
 where $\delta(\bm{\eta})$ is Dirac delta function, $I(a|A)$ denotes the integral $\int_{\bm{\eta}\neq\bm{0}} p_{o}(a|A,\bm{\eta})q_{o}(\bm{\eta})d\bm{\eta}$, and the subscript $C$ in the integral expressions means that the integral area is a small neighbourhood of the center $C$. Now we have $\int_{\tilde{B}}q'_{o}(\bm{\eta})d\bm{\eta}=1$. Then we can construct a model $\{p(a|A,\bm{\eta}),\rho(\bm{\eta}),q(\bm{\eta})\}$ by  $\rho(\bm{\eta})=\frac{1}{2}(\bm{I}+\bm{\eta}\cdot\bm{\sigma})$, $q(\bm{\eta})=q_{o}'(\bm{\eta})$, $p(a|A,\bm{\eta})=p_{o}'(a|A,\bm{\eta})$. Note that if $\int_{C}q_{o}'(\bm{0})d\bm{\eta}=0$ in Eq. (9), we let $p_{o}'(a|A,\bm{0})=\frac{1}{2}$. This model satisfies equations (2) (3), thus is an LHS model for the state.\qed\\
\indent Now we explain this geometric model. Eq. (7) sets aside the $\bm{I}$ part of every $\tilde{\rho}_{A}^{a}$ in Eq. (2) and just considers the shrinked Bloch vectors $\bm{s}_{A}^{a}$. Eq. (6) is similar to Eq. (3) except that the distribution $q(\bm{\eta})$ is not normalized. Geometrically, quantity $\mathbb{S}$ describes the quantity of vectors needed to generate a steering figure under probability conditions Eq. (6). When $\mathbb{S}\leq 1$ for a g-model, we use method given in Eq. (9) to get a normalized $q'(\bm{\eta})$, then the g-model can be directly mapped to a LHS model as showed formerly. If $\mathbb{S}_{o}>1$ for a steering figure, it is impossible to construct an LHS model for its corresponding measurement assemblages.\\
 \indent If we set aside the normalization condition $\int q(\lambda)d\lambda=1$ in Eq. (3) for LHS model $\{p(a|A,\lambda),\rho(\lambda),q(\lambda)\}$, we can directly transform a g-model $\{q(\bm{\eta})$,$p(a|A,\bm{\eta})\}$ and such a modified state model into each other, by taking $\bm{b}_{\lambda_{\bm{\eta}}}=\bm{\eta}$, $\rho(\lambda_{\bm{\eta}})=\frac{1}{2}(\bm{I}+\bm{b}_{\lambda_{\bm{\eta}}}\cdot\bm{\sigma})$, and relations in Eq. (8). We call such model the modified local hidden state (MLHS) model. Notice that a g-model may be transformed into different MLHS models, since there may be more than one $\lambda_{\bm{\eta}}$ with different distributions $\{q(\lambda_{\bm{\eta}}),p(a|A,\lambda_{\bm{\eta}})\}$ for some $\bm{\eta}$. When $\mathbb{S}\leq1$ for a g-model, the corresponding MLHS models can be changed into a valid LHS model with a method similar to Eq. (9). We will denote the MLHS models with $\mathbb{H}$ hereinafter.\\
\indent Now we classify the g-models and MLHS models into simpler equivalent groups which are one-to-one corresponded. Note that all the proves and details of the following theorems are in appendix A-D.\\
\indent $\bf{Theorem\ 1.}$ Any g-model in $\tilde{B}$ can be transformed into an equivalent g-model $\{\hat{q}(\bm{\xi}),\hat{p}(a|A,\bm{\xi})\}$ in area $N_{C}=N\cup \{C\}$ with equal quantity $\mathbb{S}$, where $N$ is the surface of $\tilde{B}$ and $C$ the center, $\bm{\xi}$ are unit vectors or zero vector.\\
\indent We call the latter g-model extreme g-model and denote it with $\mathcal{G}_{E}$.\\
\indent $\bf{Theorem\ 2.}$ Any MLHS model $\{p(a|A,\lambda),q(\lambda),\rho(\lambda)\}$ can be transformed into an equivalent one written as $\{p''(a|A,\bm{\upsilon}),q''(\bm{\upsilon}),\rho(\bm{\upsilon})\}$, where $\{\rho(\bm{\upsilon})\}$ contains only pure states and maximally mixed state. The latter MLHS models can be one-to-one mapped to extreme g-models.\\
\indent Theorem 1,2 also show that, when we try to construct concrete local models for two-qubit states by geometric methods or programming, we just need to consider a type of simpler g-models (MLHS models). We call the MLHS model which is one-to-one mapped to an extreme g-model the corresponding MLHS model of the g-model, and vice versa. The MLHS model $\mathbb{H}_{o}$ which corresponds to $\mathcal{G}_{Eo}$ is defined to be the optimal MLHS model for the measurement assemblage, since there is no LHS model for the assemblage if $\mathbb{H}_{o}$ is not an LHS model ($\mathbb{S}_{o} >1$). When $\mathbb{S}_{o}\leq 1$, $\mathbb{H}_{o}$ can be changed into the optimal LHS model by adding maximal mixed state $\bm{I}$ and normalizing its distributions $\{p''_{o}(a|A,\bm{\upsilon}),q''_{o}(\bm{\upsilon})\}$ using the method in Eq. (9). Such optimal LHS model has the largest proportion of $\bm{I}$ among all the LHS models of the assemblage.
\section{steerability quantification and model construction}
  \indent The ideas to quantify steerability have been discussed in several works before \cite{b10,b18}. In Ref. \cite{b10}, the quantification is based on the minimum amount of genuine steerable resources required to reconstruct an assemblage. Here we also propose a method of steerability quantification. We take quantity $\mathbb{S}$ as the local resource required for an MLHS model to reconstruct the assemblage. Then, quantity $\mathbb{S}_{o}$ indicates the minimum required local resources. We use $\mathbb{S}_{o}$ as the quantification of steerability of the steering assemblage. It can be shown that, when $\mathfrak{M}_{A}$ is the set of all POVMs, $\mathbb{S}_{o}$ is unchanged under local unitary (LU) operations and does not increase under local operations and shared randomness (LOSR), see appendix F.\\
   \indent Now we consider construction of g-models under state mixing.\\
\indent $\bf{Theorem\ 3.}$ A state $\rho$ obtained by $\rho=\sum_{i}c(i)\tau_{i}$ has a MLHS model $\mathbb{H}_{\rho}$ under a measurement set $\mathfrak{M}_{A}$ if every $\tau_{i}$ has a MLHS model $\mathbb{H}_{i}$ under this set, where $\tau_{i}$ are some two-qubit states, real number $c(i)\geq 0$ and $\sum_{i}c(i)=1$. \\
\indent It's known that the set of LHS is a convex set, the mixed state of unsteerable states is unsteerable. Theorem 3 gives a stronger method that may obtain unsteerable two-qubit states by mixing states with MLHS models. To do so we just need to construct $\mathbb{H}_{\rho}$ and calculate quantity $\mathbb{S}_{\rho}$ of its corresponding g-model.\\
\indent A natural question arises as, do all the two-qubit states have an MLHS model for the steered side (Bob) under arbitrary measurement set of steering side (Alice)? Using theorem 3 we demonstrate that the answer is affirmative.\\
\indent $\bf{Theorem\ 4.}$ The MLHS models exist for the set of states $\{\phi(c)\}$ under a measurement set $\mathfrak{M}_{A}$, if and only if MLHS model exists for some $\phi(c_{0})$ under $\mathfrak{M}_{A}$, where
\begin{equation}
\phi(c)=c\rho+(1-c)\rho_{A}\otimes\bm{I}_{B},
\end{equation}
where $\rho$ is an arbitrary two-qubit state, $c\in[0,1]$, $c_{0}\in(0,1]$ and $\rho_{A}=\mathrm{Tr}_{B}(\rho)$.\\
\indent Every two-qubit state $\rho$ can be written as
\begin{equation}
\rho=c_{1}\pi_{\bm{a}}+c_{2}\pi_{\bm{b}}+c_{3}\pi_{\bm{T}},
\end{equation}
where coefficients $c_{i}$ $(i=1,2,3)$ are nonnegative real numbers and $\sum_{i}c_{i}=1$. States $\pi_{\bm{a}}=\frac{1}{4}(\bm{I}+\bm{a}\cdot\bm{\sigma}\otimes\bm{I}_{B})$, $\pi_{\bm{b}}=\frac{1}{4}(\bm{I}+\bm{I}_{A}\otimes\bm{b}\cdot\bm{\sigma})$, $\pi_{\bm{T}}=\frac{1}{4}(\bm{I}+\sum_{i,j=1}^{3}T_{ij}\sigma_{i}\otimes\sigma_{j})$, where $\sigma_{i,j}$ are pauli matrices, $\bm{a}$ and $\bm{b}$ are two vectors satisfying $|\bm{a}|\leq1$ and $|\bm{b}|\leq1$, $T_{ij}$ are real and $|T_{ij}|\leq 1$.\\
\indent Since states $\pi_{\bm{a}}$ and $\pi_{\bm{b}}$ are separable states, LHS models exist for them under any measurement set. $\pi_{\bm{T}}$ are states which satisfy $G=\begin{pmatrix}  1&\bm{0}\\ \bm{0}&\bm{T} \end{pmatrix}$. Such states are called T states \cite{b11}. We have proved that steerability is unchanged under local unitary (LU) operations in appendix, so we just need to find g-models for a subset of T states, which have diagonal $\bm{T}$ matrices without loss of generality \cite{b11}. Let $\pi_{D}$ denote such states, which can be written as
\begin{equation}
\pi_{D}=\frac{1}{4}(\bm{I}\otimes \bm{I}+\sum_{i=1}^{3}T_{ii}\sigma_{i}\otimes\sigma_{i}).
\end{equation}
\indent This subset of states are called Bell diagonal states since they can be written as combinations of four Bell states $\rho_{Bi}$ $(i=1,2,3,4)$. As the states $\rho(p)=p\rho_{Bi}+(1-p)\bm{I}/2$ are separable states (thus have LHS models) when $p\leq\frac{1}{3}$, the four Bell states have MLHS models under any measurement set according to theorem 4. Then, from theorem 3 we know that all $\pi_{D}$ (thus all $\pi_{\bm{T}}$) have MLHS models.\\
\indent As three types of states in Eq. (11) have MLHS models under an arbitrary $\mathfrak{M}_{A}$, using theorem 3 we obtain that all two-qubit states have MLHS models under an arbitrary $\mathfrak{M}_{A}$. This result also supports the idea to quantify steerability of two-qubit states using quantity $\mathbb{S}_{o}$.
\section{Specific examples}
\indent Now we give some applications of Theorem 3. The examples will be given using extreme g-model under continuous sets of all projective measurements. All measurements $A$ are assumed to be projective measurement performed by Alice in the following cases.\\
\indent\emph{1.Werner states}\\
  \indent Two-qubit Werner states \cite{b12} can be written as
\begin{equation}
W(p)=p|\psi\rangle \langle\psi|+(1-p)\bm{I}/4,
\end{equation}
where $|\psi\rangle \langle\psi|$ is the singlet state and $\bm{I}$ is the identity. According to the existing optimal LHS model for some Werner states \cite{b2,b12}, we obtain that the optimal g-model for a two-qubit singlet state $|\psi\rangle$ (including Werner state)  under projective measurements is
  \begin{align}
p_{Eo}(a|A,&\bm{\xi})=\left \{
\begin{aligned}
1, &\qquad |\bm{\xi}|=1, \bm{\xi}\cdot \bm{s}_{A}^{a}\geq 0,\\
1/2, &\qquad |\bm{\xi}|=0,\\
0, &\qquad otherwise,
\end{aligned}
\right.\nonumber\\
&q_{Eo}(\bm{\xi})=\left \{
\begin{aligned}
\frac{1}{2\pi}, &\qquad |\bm{\xi}|=1,\\
0, &\qquad |\bm{\xi}|=0,
\end{aligned}
\right.
\end{align}
where subscript $Eo$ indicates that it is an optimal extreme g-model. The optimality is proved in our another work \cite{b13}.\\
\indent For all Bell diagonal states, $p(a|A)=\frac{1}{2}$. In the proof of theorem 3 we demonstrated that, for such states, if we construct an MLHS model $\mathbb{H}_{\rho}$ using the way we proposed, the quantity $\mathbb{S}_{\rho}$ satisfies $\mathbb{S}_{\rho}=\sum_{i}c(i)\mathbb{S}_{i}$. As the quantity $\mathbb{S}_{o}$ is 2 for $|\psi\rangle$ and 0 for $\bm{I}$, we obtain that Werner state is unsteerable if $p\leq \frac{1}{2}$.\\
\indent \emph{2.Mixed states of 2D T states and singlet state}\\
\indent  Note that we just need to consider states $\pi_{D}$ with diagonal $\bm{T}$ matrix. 2D T states are $\pi_{D}$ that $T_{ii}$ vanishes for an arbitrary i, their steering figures are 2D ellipses centered at the center of $\tilde{B}$.\\
\indent In another work \cite{b13} we give an optimal g-model for 2D T states and provided a way of calculating quantity $\mathbb{S}_{o}$ under this model. We obtain that quantity $\mathbb{S}_{o}$ equals half of the circumference of the steering ellipse for 2D T states. \cite{b13}\\
\indent Here we illustrate the case with a state $\rho$ satisfying $T_{11}=T_{22}=-1/2$, $T_{33}=0$, whose steering ellipse is a circle with radius $1/4$ and circumference $\pi/2$.\\
\indent In the proof of theorem 3 in appendix we also prove that the mixture of T states are unsteerable if the mixture of their steering quantity $\mathbb{S}$ does not exceed $1$. The quantity $\mathbb{S}_{o}$ for $\rho$ is $\pi/4$, so the state $\rho(p)$ generated by
\begin{equation}
\rho(p)=p|\psi\rangle \langle\psi|+(1-p)\rho,
\end{equation}
is unsteerable if $p\leq \frac{1-\pi/4}{2-\pi/4}$. Such unsteerable states are outside the convex tetrahedron of separable $\pi_{D}$ (except for $p=0$ case), thus they are states that are entangled but unsteerable under all projective measurements. This result can be examined using the criteria proposed in some former works about the steerability of T states \cite {b15,b16}.\\
\indent \emph{3.Mixture of states $\tau_{i}$ with different} $p_{i}(a|A)$\\
\indent Let $\tau_{i}$ be some two-qubit states whose g-models are constructed and with different probability $p_{i}(a|A)$. We can construct a g-model for $\rho$ by mixing the g-models of $\tau_{i}$ as showed in proof of Theorem 3 in appendix. The corresponding $\mathbb{S}_{\rho}$ satisfies $\mathbb{S}_{\rho}\geq\sum_{i}c(i)\mathbb{S}_{i}$ and can obtained by simple calculation. Especially when $q_{i}(\bm{0})=0$, we can let $\mathbb{S}_{\rho}=\frac{\sum_{i}c(i)p_{i}(a|A)\mathbb{S}_{i}}{p_{\rho}(a|A)}$. In Ref. \cite{b4,b16}, LHS models for some two-qubit states with nonuniform $p(a|A)$ are given. By transforming them into g-models and mixing them with other states, we can obtain more untrivial unsteerable states.\\
\indent Since any POVM can be converted into an equivalent ``fine grained'' one, with each operator $A_{a}$ written as a shrinked projector as $A_{a}=c_{a}P_{a}$ $(0<c_{A}\leq1)$ \cite{b17}, a question is raised as, can we generalize the result for projectors into POVM operators by shrinking the conditioned probabilities of MLHS models we constructed for projectors? In appendix E we show that the models constructed by this approach do not satisfy Eqs. (6) and (7) simultaneously. So constructing MLHS models under general POVM would be a more challenging task.\\
\section{conclusion}
\indent  We construct a geometric characterization of LHS models for the steering assemblages of two-qubit states, by introducing geometric model (g-model) and modified LHS model. Our work provides access to constructing concrete MLHS models for two-qubit states. More precisely, we can start with the T states. Then we can generalize the result for T states into more states. Also we may obtain some results about asymmetric steering. In another work \cite{b13} we further explore the projects above. Another interesting project is to generalize the model for generalized bipartite states of higher dimensions. Although we don't have a geometric characterization for general higher dimensional bipartite states, we might get some results for some highly symmetric states, such as isotropic states. This is left for future study.
\section*{acknowledgements}
\indent This work was supported by the National Natural Science Foundation of China (Grant No. 11275182), National Key R \& D Program (Grant No.2016YFA0301700), the Strategic Priority Research Program of the Chinese Academy of Sciences (Grant No. XDB01030300).

\section*{appendix a: Construction of the extreme g-model $\mathcal{G}_{E}$}
\indent Now we show that every g-model $\{{q}(\bm{\eta}),{p}(a|A,\bm{\eta})\}$ in the main text can be transformed into a g-model $\{\hat{q}(\bm{\xi}),\hat{p}(a|A,\bm{\xi})\}$ with equal quantity $\mathbb{S}$, where $\bm{\xi}$ are unit vectors on surface $N$ or zero vector at center $C$ of the sphere $\tilde{B}$. We denote the latter g-model as extreme g-model $\mathcal{G}_{E}$.\\
\indent Let $d\bm{\xi}$ denote the infinitesimal area on the surface $N$, corresponding to the unit vector $\bm{\xi}$. Also we label each $\bm{\eta}$ with $\bm{\xi}$ and $k\in \mathbb{R}, k\in[0,1]$, then every $\bm{\eta}$ can also be written as $\bm{\eta_\xi}^{k}$ for some $\bm{\eta}$ and $k$, which satisfies $\bm{\eta_\xi}^{k}=k\cdot\bm{\xi}$. Then we let\\
\begin{align}
\hat{q}(\bm{\xi}&)=\int_{0}^{1} q(\bm{\eta}_{\bm{\xi}}^{k})|\bm{\eta}_{\bm{\xi}}^{k}|f(\bm{\eta}_{\bm{\xi}}^{k})dk, &\bm{\xi}\neq\bm{0},\tag{s.1}\\
\hat{p}(\bm{\xi})&=\int_{\tilde{B}} q(\bm{\eta})(1-|\bm{\eta}|)d\bm{\eta}, &\bm{\xi}=\bm{0};\tag{s.2}
\end{align}
and
\begin{align}
\hat{p}(a|A,\bm{\xi}&)=\frac {\int_{0}^{1} p(a|A,\bm{\eta}_{\bm{\xi}}^{k})q(\bm{\eta}_{\bm{\xi}}^{k})|\bm{\eta}_{\bm{\xi}}^{k}| f(\bm{\eta}_{\bm{\xi}}^{k})dk}{\hat{q}(\bm{\xi})}, &\bm{\xi}\neq\bm{0},\tag{s.3}\\
\hat{p}(a|A,\bm{\xi})&=\frac {\int_{\tilde{B}} p(a|A,\bm{\eta})q(\bm{\eta})(1-|\bm{\eta}|)d\bm{\eta}}{\hat{p}(\bm{\xi})}, &\bm{\xi}=\bm{0},\tag{s.4}
\end{align}
where $f(\bm{\eta}_{\bm{\xi}}^{k})=\frac{|\bm{\eta}_{\bm{\xi}}^{k}|^{2}}{|\bm{\xi}|^{2}}=|\bm{\eta}_{\bm{\xi}}^{k}|^{2}$, and the distributions without hats are of the original g-model. Note that the probability of $\bm{\xi}$ at $C$ should be a discrete $\hat{p}(\bm{0})$, but for convenience we will also write it as $\hat{q}(\bm{0})$ in some following equations, which satisfies $\int_{C}\hat{q}(\bm{0})d\bm{\xi}=\hat{p}(\bm{0})$ and $\int_{C}\hat{p}(a|A,\bm{0})\hat{q}(\bm{0})d\bm{\xi}=\hat{p}(a|A,\bm{0})\hat{p}(\bm{0})$.\\
\indent Here we explain the construction of the extreme g-model. In Eqs. (s.1) and (s.2) we divided $q(\bm{\eta_\xi}^{k})$ into to parts, one with proportion $|\bm{\eta}_{\xi}|$ and the other with proportion $1-|\bm{\eta_\xi}^{k}|$. Then we put the first part into the distribution of the unit vector $\bm{\xi}$ and the second part into the distribution of $\bm{0}$. Note that we are also converting the distribution $q(\bm{\eta})$ in $\tilde{B}$ into a $q(\bm{\xi})$ on $N$ and $C$, that is why we have the term $f(\bm{\eta}_{\bm{\xi}}^{k})$, which satisfies $d\bm{\eta}_{\bm{\xi}}^{k}=f(\bm{\eta}_{\bm{\xi}}^{k})dkd\bm{\xi}$, in Eqs. (s.1) and (s.3).\\
\indent Now we examine if $\{\hat{q}(\bm{\xi}),\hat{p}(a|A,\bm{\xi})\}$ is a g-model for the steering figure. From Eqs. (s.1-s.4), we can obtain that for all $\bm{\xi}$, there are
\begin{align}
&\quad \sum_{a}\hat{p}(a|A,\bm{\xi})=1,\tag{s.5}
\end{align}
\begin{align}
&\int_{N_{C}} \hat{q}(\bm{\xi})\hat{p}(a|A,\bm{\xi})d\bm{\xi}\nonumber\\
=\int_{\tilde{B}} p(a|&A,\bm{\eta})q(\bm{\eta})d\bm{\eta}=p(a|A)\cdot\int_{\tilde{B}} q(\bm{\eta})d\bm{\eta}\nonumber\\=p(a|A)&\int_{N_{C}} \hat{q}(\bm{\xi})d\bm{\xi},\tag{s.6}
\end{align}
where $N_{C}=N\cup \{C\}$, which is the combined area of the surface and the center. And we have
\begin{align}
&\int_{N_{C}} \hat{q}(\bm{\xi})\hat{p}(a|A,\bm{\xi})\bm{\xi}d\bm{\xi}\nonumber\\
= \int& _{\tilde{B}} p(a|A,\bm{\eta})q(\bm{\eta})\bm{\eta}d\bm{\eta}=\bm{s}_{A}^{a},\tag{s.7}
\end{align}
which altogether indicate that $\{\hat{p}(a|A,\bm{\xi}),\hat{q}(\bm{\xi})\}$ is also a g-model for the same steering figure.\\
\indent The quantity $\mathbb{S}$ for the extreme g-model is
\begin{equation}
\mathbb{S}=\int_{N_{C}} \hat{q}(\bm{\xi})d\bm{\xi}=\int_{\tilde{B}}q(\bm{\eta})d\bm{\eta},\tag{s.8}
\end{equation}
which shows that quantity $\mathbb{S}$ doesn't change when the original g-model is transformed into an extreme one.\\
\indent Now that any original g-model can always be transformed into an extreme g-model with same quantity $\mathbb{S}$, we can say that all g-models which can be transformed into the same extreme g-model are equivalent to this extreme g-model, and we just need to consider the extreme case when constructing a concrete model.\\

\section*{appendix b: Construction of equivalent MLHS models}
\indent For every MLHS model $\{q(\lambda),p(a|A,\lambda),\rho(\lambda)\}$, where
\begin{equation}
\rho(\lambda)=\frac{1}{2}(\bm{I}+\bm{b}_{\lambda}\cdot\bm{\sigma}),\tag{s.9}
\end{equation}
we obtain its g-model $\{q(\bm{\eta}),p(a|A,\bm{\eta})\}$ using (8) in the main text. Then we transform the original g-model into an extreme one, written $\{\hat{q}(\bm{\xi}),\hat{p}(a|A,\bm{\xi})\}$.\\
\indent Now we can construct an MLHS model $\{q''(\bm{\upsilon}),p''(a|A,\bm{\upsilon}),\rho(\bm{\upsilon})\}$ where
\begin{equation}
\rho(\bm{\upsilon})=\frac{1}{2}(\bm{I}+\bm{\upsilon}\cdot\bm{\sigma}),\tag{s.10}
\end{equation}
and $\bm{\upsilon}$ are unit vectors or zero vectors, by letting
\begin{align}
q''(\bm{\upsilon})&=\hat{q}(\bm{\xi})_{|\bm{\upsilon}=\bm{\xi}},\nonumber\\
p''(a|A,\bm{\upsilon})&=\hat{p}(a|A,\bm{\xi})_{|\bm{\upsilon}=\bm{\xi}},\tag{s.11}
\end{align}
for every $\bm{\upsilon}$ and $\bm{\xi}$. \\
\indent Simple calculations shows that the model $\{p''(a|A,\bm{\upsilon}),q''(\bm{\upsilon}),\rho(\bm{\upsilon})\}$ is also an MLHS model for the state. This MLHS model, obtained by (s.11), is one-to-one mapped to an extreme g-model. We call the MLHS model which one-to-one mapped to an extreme g-model the corresponding MLHS model of the g-model, and vice versa.\\
\indent Since the g-modes of the original MLHS model $\{q(\lambda),p(a|A,\lambda),\rho(\lambda)\}$ is equivalent to the extreme g-model of the transformed MLHS model $\{q''(\bm{\upsilon}),p''(a|A,\bm{\upsilon}),\rho(\bm{\upsilon})\}$, we say these two MLHS models have the same steerability and are equivalent. We can use the latter MLHS model to represent all MLHS models in the same equivalent group.\\

\section*{appendix c: Proof of theorem 3}
\indent First we review theorem 3 in the main text.
\indent $\bf{Theorem\ 3.}$ A two qubit state $\rho$ obtained by mixture $\rho=\sum_{i}c(i)\tau_{i}$ has an MLHS model $\mathbb{H}_{\rho}$ under $\mathfrak{M}_{A}$ if every $\tau_{i}$ has an MLHS model $\mathbb{H}_{i}$ under $\mathfrak{M}_{A}$, where $\tau_{i}$ are some two-qubit states, real number $c(i)\geq 0$ and $\sum_{i}c(i)=1$. \\
\indent \emph{Proof.}   Here we construct the corresponding g-model of $\mathbb{H}_{\rho}$. Let $p_{i}(a|A)$ and $\bm{s}_{iA}^{a}$ denote the probabilities and conditioned vectors under $\mathfrak{M}_{A}$ for $\tau_{i}$. For state $\rho=\sum_{i}c(i)\tau_{i}$, there is $p_{\rho}(a|A)=\sum_{i}c(i)p_{i}(a|A)$ and $\bm{s}_{\rho A}^{a}=\sum_{i}c(i)\bm{s}_{iA}^{a}$. We construct a model $\{q_{\rho}(\bm{\eta}),p_{\rho}(a|A,\bm{\eta})\}$ satisfying
\begin{align}
q_{\rho}(\bm{\eta})&=\sum_{i}c(i)q_{i}(\bm{\eta}),\nonumber\\
p_{\rho}(a|A,\bm{\eta})=&\frac{\sum_{i}c(i)q_{i}(\bm{\eta})p_{i}(a|A,\bm{\eta})}{q_{\rho}(\bm{\eta})},\tag{s.12}
 \end{align}
  for every $\bm{\eta}$, where $\{q_{i}(\bm{\eta}),p_{i}(a|A,\bm{\eta})\}$ is the g-model (not necessarily optimal) for $\tau_{i}$ under $\mathfrak{M}_{A}$. \\
  \indent Let $\mathbb{S}_{I}$ denote the sum $\sum_{i}c(i)\mathbb{S}_{i}$.\\
  \indent Similar to (9) in the main text, we construct the modified nonnegative distributions $\{q_{\rho}'(\bm{\eta}),p_{\rho}'(a|A,\bm{\eta})\}$ satisfying
\begin{align}
&q_{\rho}'(\bm{\eta})=(\mathbb{S}_{\rho}-\mathbb{S}_{I})\cdot\delta(\bm{\eta})+q_{\rho}(\bm{\eta}),\nonumber\\
p_{\rho}'(a|A,&\bm{\eta})=\left \{
\begin{aligned}
&\frac{p_{\rho}(a|A)\mathbb{S}_{\rho}-L(a|A)}{\int_{C}q_{\rho}'(\bm{0})d\bm{\eta}}, &\quad \bm{\eta}=\bm{0},\\
&p_{\rho}(a|A,\bm{\eta}), &\quad otherwise,
\end{aligned}
\right.\tag{s.13}
\end{align}
 where $\delta(\bm{\eta})$ is Dirac delta function, $L(a|A)$ denotes the integral $\sum_{i}c(i)\int_{\bm{\eta}\neq\bm{0}} p_{i}(a|A,\bm{\eta})q_{i}(\bm{\eta})d\bm{\eta}$, $\mathbb{S}_{\rho}$ is a value which keeps $q_{\rho}'(\bm{\eta})\geq0$ and $p_{\rho}(a|A)\mathbb{S}_{\rho}-L(a|A)\geq0$ for all $a$ and $A$. Since $L(a|A)\leq \sum_{i}c(i)\mathbb{S}_{i}p_{i}(a|A)\leq\mathbb{S}_{m}p_{\rho}(a|A)$, where $\mathbb{S}_{m}=$max$_{i}\{\mathbb{S}_{i}\}$, we can always find an $\mathbb{S}_{\rho}$ satisfying $\mathbb{S}_{\rho}\leq \mathbb{S}_{m}$, which means the quantity $\mathbb{S}_{\rho}$ would not be larger than the largest $\mathbb{S}_{i}$ if it is well chosen.\\
 \indent From (s.13) there is
  \begin{align}
 0\leq p_{\rho}'(a|A,\bm{\eta})\leq1,\nonumber\\ \sum_{a} p_{\rho}'(a|A,\bm{\eta})=1,\tag{s.14}
  \end{align}
\indent and
\begin{align}
&\qquad \int _{\tilde{B}} p_{\rho}'(a|A,\bm{\eta})q_{\rho}'(\bm{\eta})\bm{\eta}d\bm{\eta}\nonumber\\&=\sum_{i}c(i)\int _{\tilde{B}} p_{i}(a|A,\bm{\eta})q_{i}(\bm{\eta})\bm{\eta}d\bm{\eta}\nonumber\\&=\sum_{i}c(i)\bm{s}_{iA}^{a}=\bm{s}_{\rho A}^{a}.\tag{s.15}
\end{align}
Also
\begin{align}
 &\qquad \int_{\tilde{B}} p_{\rho}'(a|A,\bm{\eta})q_{\rho}'(\bm{\eta})d\bm{\eta}\nonumber\\&=p_{\rho}(a|A)\mathbb{S}_{\rho}-L(a|A)+L(a|A)\nonumber\\&=p_{\rho}(a|A)\mathbb{S}_{\rho},\tag{s.16}
  \end{align}
\indent Therefore $\{q_{\rho}'(\bm{\eta}),p_{\rho}'(a|A,\bm{\eta})\}$ is a g-model for $\rho$, and its quantity $\mathbb{S}'$ is
  \begin{align}
  \mathbb{S}'&=\int_{\tilde{B}} q_{\rho}'(\bm{\eta})d\bm{\eta}\nonumber\\=\sum_{i}c(i)\int_{\tilde{B}} &q_{i}(\bm{\eta})d\bm{\eta}+(\mathbb{S}_{\rho}-\mathbb{S}_{I})=\mathbb{S}_{\rho}.\tag{s.17}
  \end{align}
\indent Specially, when all $\tau_{i}$ are T states, their $p_{\rho}(a|A)$ are constant $\frac{1}{2}$ for all $a$ and $A$. If $L(a|A)$ are also constant $\frac{1}{2}$ for them, we can let $\mathbb{S}_{\rho}=\mathbb{S}_{I}$ and $q_{\rho}'(\bm{\eta})=q_{\rho}(\bm{\eta})$, $p_{\rho}'(a|A,\bm{\eta})=p_{\rho}(a|A,\bm{\eta})$. Therefore the mixture of T states are unsteerable if the mixture of their quantities $\mathbb{S}$ does not exceed 1.\qed\\

\section*{appendix d: Proof of theorem 4}
\indent First we review theorem 4 in the main text.\\
\indent $\bf{Theorem}\ 4.$ \ The MLHS models exist for the set of states $\{\phi(c)\}$ under a measurement set $\mathfrak{M}_{A}$, if and only if MLHS model exists for some $\phi(c_{0})$ under $\mathfrak{M}_{A}$, where
\begin{equation}
\phi(c)=c\rho+(1-c)\rho_{A}\otimes\bm{I}_{B},
\end{equation}
where $\rho$ is an arbitrary two-qubit state, $c\in[0,1]$, $c_{0}\in(0,1]$ and $\rho_{A}=Tr_{B}(\rho)$.\\
\indent \emph{Proof} \ The necessity is obvious so we prove the sufficiency. Note first that MLHS model for a two qubit state is equivalent to g-model for its ellipsoid. We denote the g-model for $\phi(c_{0})$ under $\mathfrak{M}_{A}$ as $\{q_{c_{0}}(\bm{\eta}),p_{c_{0}}(a|A,\bm{\eta})\}$. For any state $\phi(c)$ in the set, its probability $p_{c}(a|A)$ upon performing measurement operator $A_{a}$ is $p_{c}(a|A)=Tr(A_{a}\rho_{A})$, which is independent of $c$. Its conditioned vectors $\bm{s}_{cA}^{a}$ satisfy $\bm{s}_{cA}^{a}=\frac{c}{c_{0}}\cdot\bm{s}_{c_{0}A}^{a}$.\\
\indent Then for every $\phi(c)$ $(1\geq c\geq0)$, distributions
 \begin{align}
q_{c}(\bm{\eta})&=\frac{c}{c_{0}}q_{c_{0}}(\bm{\eta}),\nonumber\\
p_{c}(a|A,\bm{\eta})&=p_{c0}(a|A,\bm{\eta}),  \tag{s.19}
\end{align}
 satisfies equations (6) and (7) in the main text. Therefore $\{q_{c}(\bm{\xi}),p_{c}(a|A,\bm{\xi})\}$ is a g-model for $\phi(c)$, with steering quantity $\mathbb{S}_{c}=\frac{c}{c_{0}}\cdot \mathbb{S}_{c_{0}}$. Now that there is a g-model for the steering ellipsoid of every $\phi(c)$, there is an MLHS model for every $\phi(c)$ in the set under $\mathfrak{M}_{A}$.\qed\\

\section*{appendix e: The difficulty of generating MLHS model into POVM}
\indent Consider a POVM set $\{E_{i}\}$. Since any POVM can be converted into an equivalent "fine grained" one, with each operator written as a shrinked projector \cite{b17}, we let every measurement operater $E_{i}$ be a shrinked projector of a rank 1 projector $P_{i}$ without loss of generality. Then we have $E_{i}=c_{i}P_{i}$ , where $c_{i}$ are real numbers satisfying $0\leq c_{i}\leq1$ and $\sum_{i}c_{i}=2$ for two qubit states. Note that although projectors $P_{i}$ satisfy $\sum_{i}P_{i}=\bm{I}$, they may come from different sets of projective measurements and be non-orthogonal.\\
\indent Now a question arises as, can we construct an MLHS model for general POVM operators $E_{i}$ by multiplying $c_{i}$ to the conditioned probability $p(a|A,\bm{\upsilon})$ under projectors $P_{i}$ while keeping $q(\bm{\upsilon})$ unchanged? We'll show that this can not be done.\\
\indent The probability of a state $\rho$ obtaining result $i$ under POVM operator $E_{i}$ is $p_{E}(i|E)=c_{i}p(i|P_{i})$ and the unnormalized conditioned state is $\tilde{\rho_{E_{i}}}=c_{i}\tilde{\rho_{i}}$, where $p(i|P_{i})$ and $\tilde{\rho_{i}}$ are the conditioned probability and unnormalized conditioned state under projector $P_{i}$.\\
\indent We denote the MLHS model of the state under projector $P_{i}$ as $\{q''(\bm{\upsilon}),\rho_{\upsilon},p''(i|P_{i},\bm{\upsilon})\}$. Now we let $p_{E}''(i|E,\bm{\upsilon})=c_{i}p''(i|P_{i},\bm{\upsilon})$ and check if model model $\{q''(\bm{\upsilon}),\rho_{\upsilon},p_{E}''(i|E,\bm{\upsilon})\}$ is the MLHS model for the state under POVM operators $E_{i}$. \\
\indent First there is
\begin{align}
\int q''(\bm{\upsilon})p_{E}''(i|E,\bm{\upsilon})&d\bm{\upsilon}=c_{i}\int q''(\bm{\upsilon})p''(i|P_{i},\bm{\upsilon})d\bm{\upsilon}\nonumber\\&=p_{E}(i|E).\tag{s.20}
\end{align}
\indent For every unnormalized conditioned states there is
\begin{align}
\int p_{E}''(i|E,\bm{\upsilon}) &q''(\bm{\upsilon}) \rho_{\upsilon}d\bm{\upsilon}\nonumber\\=c_{i}\int p''(i|P_{i},\bm{\upsilon}) q''(\bm{\upsilon})& \rho_{\upsilon}d\bm{\upsilon}=\tilde{\rho_{E_{i}}}.\tag{s.21}
\end{align}
\indent Equations (s.20) and (s.21) show that directly generating the GHLS models for projectors into models for POVM operators can reproduce the conditioned probabilities and conditioned states of Alice and Bob.\\
\indent However, another condition for MLHS model is
\begin{equation}
\sum_{i} p_{E}''(i|E,\bm{\upsilon})=1.\tag{s.22}
\end{equation}
 For the model we construct, there is
\begin{align}
\sum_{i} p_{E}''(i|E,\bm{\upsilon})=\sum_{i}c_{i} p''(i|P_{i},\bm{\upsilon}).\tag{s.23}
\end{align}
\indent If $P_{i}$ are from the same measurement set, then $\sum_{i}c_{i} p''(i|P_{i},\bm{\upsilon})=1$. But for an arbitrary $\rho$ and non-orthogonal $P_{i}$, it's not always the case.\\
\indent For example, a g-model (denoted with $H$) for a singlet state $|\psi\rangle$ under projective measurements is
\begin{align}
p_{o}''(&i|P_{i},\bm{\upsilon})=\left \{
\begin{aligned}
1, &\qquad \bm{\upsilon}\cdot \bm{b}_{P_{i}}\geq 0, |\bm{\upsilon}|=1,\\
1/2, &\qquad|\bm{\upsilon}|=0,\\
0, &\qquad otherwise,
\end{aligned}
\right.\nonumber\\
&q_{o}''(\bm{\upsilon})=\left\{
\begin{aligned}
\frac{1}{2\pi}, &\qquad \bm{\upsilon}\neq \bm{0},\\
0, &\qquad \bm{\upsilon}= \bm{0},
\end{aligned}
\right. \tag{s.24}
\end{align}
where $\bm{\upsilon}$ are unit vectors or zero vectors, $\bm{b}_{P_{i}}$ is the Bloch vector of the conditioned state $\rho_{P_{i}}$ under $P_{i}$.\\
\indent Consider a POVM set $E_{1}=\frac{2}{3}|0\rangle\langle0|$, $E_{2}=\frac{2}{3}|\alpha\rangle\langle\alpha|$, $E_{3}=\frac{2}{3}|\beta\rangle\langle\beta|$, where $|\alpha\rangle=\frac{1}{2}|0\rangle+\frac{\sqrt{3}}{2}|1\rangle$, $|\beta\rangle=\frac{1}{2}|0\rangle-\frac{\sqrt{3}}{2}|1\rangle$ are both pure states. The only set of projectors that can realize this POVM is $\{|0\rangle\langle0|,|\alpha\rangle\langle\alpha|,|\beta\rangle\langle\beta|\}$. Let $P_{1}$, $P_{2}$, $P_{3}$ denote these three projectors. Obviously the conditioned probability $p_{E}''(i|E,\bm{\upsilon})=\frac{1}{3}\sum_{i=1}^{3}p_{o}''(i|P_{i},\bm{\upsilon})$ for $E_{i}$  doesn't satisfy condition (s.22), thus $\{q_{o}''(\bm{\upsilon}),\rho_{\upsilon},p_{E}''(i|E,\bm{\upsilon})\}$ can not be an MLHS model for singlet state under POVMs.\\
\indent Even if we adjust the coefficient $c_{i}$ for every $p''(i|P_{i},\bm{\upsilon})$ or add a normalization term to $p_{E}''(i|E,\bm{\upsilon})$, it's still hard to construct an MLHS model for all general POVM operators by projective measurements satisfying (s.20-22) simultaneously.\\

\section*{appendix f: Proof that $\mathbb{S}_{o}(\rho)$ does not change in local unitary $(LU)$ operations and does not increase in local operations and shared randomness $(LOSR)$ under the set of all POVMs.}
\indent LOSR are completely positive trace-preserving (CPTP) maps that can be written as $\sum_{k}u(k)\mathcal {E}_{k}\otimes\mathcal{F}_{k}$, where $\mathcal {E}_{k}$ and $\mathcal{F}_{k}$ are $CPTP$ maps on the linear operators of Alice and Bob's Hilbert space respectively, $u(k)$ is probability distribution \cite{b19}.\\
\indent When an $LOSR$ operation is performed on the two-qubit state $\rho$, the result state $\overline{\rho}$ is
\begin{equation}
\overline{\rho}=\sum_{k}u(k)\sum_{ij}L_{Xki}\otimes L_{Ykj}\rho L_{Xki}^{\dag}\otimes L_{Ykj}^{\dag},\tag{s.25}
\end{equation}
where $X$ denotes the local system of Alice and $Y$ denotes the local system of Bob, $L_{Xki}$ and $L_{Yki}$ are positive operators satisfying $\sum_{i}L_{Xki}L_{Xki}^{\dag}=\bm{I}_{X}$ and $\sum_{j}L_{Ykj}L_{Ykj}^{\dag}=\bm{I}_{Y}$.\\
\indent When Alice performs POVM operator $M_{A}^{a}$ on her own system of $\overline{\rho}$, the unnormalized conditioned state $\overline{\tilde{\rho}_{A}^{a}}$ of Bob's side is
\begin{align}
\overline{\tilde{\rho}_{A}^{a}}=&Tr_{X}(M_{A}^{a}\otimes \bm{I}_{B}\overline{\rho})\nonumber\\
&=\sum_{k}u(k)\tilde{\rho}_{A}^{a}(k)\tag{s.26},
\end{align}
in which
\begin{align}
\tilde{\rho}_{A}^{a}(k)&=Tr_{X}(\overline{M_{A_{k}}^{a_{k}}}\otimes \bm{I}_{B}\sum_{j}\bm{I}_{A}\otimes L_{Ykj}\rho\bm{I}_{A}^{\dag}\otimes L_{Ykj}^{\dag})\nonumber\\
&=\sum_{jk}L_{Ykj}Tr_{X}(\overline{M_{A_{k}}^{a_{k}}}\otimes \bm{I}_{B}\rho)L_{Ykj}^{\dag},\tag{s.27}
\end{align}
where $\{\overline{M_{A_{k}}^{a_{k}}}\}$ is also a set of POVM, with $\overline{M_{A_{k}}^{a_{k}}}=\sum_{i}L_{X_{ki}}^{\dag}M_{A}^{a}L_{X_{ki}}$.\\
\indent Suppose the optimal MLHS model for $\rho$ under POVM set $\{M_{A}^{a}\}$ is $\{q(\lambda),p(a|A,\lambda),\rho(\lambda)\}$, then we can construct an MLHS model for $\rho_{LOSR}$ by two steps. First we introduce MLHS model $\mathbb{H}_{k}:\{q_{k}(\lambda_{k}),p_{k}(a|A,\lambda_{k}),\rho_{k}(\lambda_{k})\}$ for each set of assemblages $\{\tilde{\rho}_{A}^{a}(k)\}$, where
\begin{align}
\rho_{k}(\lambda_{k})=&\sum_{i}L_{Yki}\rho(\lambda) L_{Yki}^{\dag},\nonumber\\
p_{k}(a|A,\lambda_{k}&)=p(a_{k}|A_{k},\lambda),\nonumber\\
q_{k}(\lambda_{k}&)=q(\lambda).\tag{s.28}
\end{align}
Notice that the quantity $\mathbb{S}_{k}$ corresponding to each $\mathbb{H}_{k}$ satisfies $\mathbb{S}_{k}=\mathbb{S}_{o}(\rho)$.\\
\indent Then, we combine these MLHS models into a new MLHS model $\mathbb{H'}:\{q'(\bm{\gamma}),p'(a|A,\bm{\gamma}),\rho'(\bm{\gamma})\}$, where $\bm{\gamma}$ is the vectors inside a unit sphere (satisfying $|\bm{\gamma}|\leq 1$), and every $\rho'(\bm{\gamma})$ satisfies
\begin{align}
\rho'(\bm{\gamma})=\frac{1}{2}(\bm{I}+\bm{\gamma}\cdot\bm{\sigma}).\tag{s.29}
\end{align}
We construct this new MLHS model by letting
\begin{align}
&q'(\bm{\gamma})=\sum_{\{k|\rho_{k}(\lambda_{k})=\rho'(\bm{\gamma})\}}u(k)q_{k}(\lambda_{k}),\nonumber\\
p'(a|A,\bm{\gamma})=&\frac{\sum_{\{k|\rho(\lambda_{k})=\rho'(\bm{\gamma})\}}u(k)p_{k}(a|A,\lambda_{k})q_{k}(\lambda_{k})}{q'(\bm{\gamma})}.\tag{s.30}
\end{align}
Using the result in the proof of theorem $3$, we can obtain that the MLHS model $\mathbb{H}'$ can reconstruct the assemblages $\overline{\tilde{\rho}_{A}^{a}}$, also, its corresponding quantity $\mathbb{S}'\leq$ max$_{k}$$\{\mathbb{S}_{k}\}=\mathbb{S}_{o}(\rho)$.\\
\indent Therefore we have
\begin{equation}
\mathbb{S}_{o}(\overline{\rho})\leq \mathbb{S}'\leq\mathbb{S}_{o}(\rho),\tag{s.31}
\end{equation}
which proves that the quantity $\mathbb{S}_{o}$ does not increase under $LOSR$.\\
\indent Now we consider the case that the LOSR they perform is an LU operation, written $U_{X}\otimes U_{Y}$. LU operations are reversible, and the inverse operation of an LU operation is also an LU operation. Let $\overline{\rho}_{LU}$ denote the state obtained by performing the LU operation $U_{X}\otimes U_{Y}$ on $\rho$. By applying $U_{X}\otimes U_{Y}$ on $\rho$ and applying $U_{X}^{-1}\otimes U_{Y}^{-1}$ on $\overline{\rho}_{LU}$ we can transform the two states into each other, and according to former result we have $\mathbb{S}_{o}(\overline{\rho})\leq\mathbb{S}_{o}(\rho)$ and $\mathbb{S}_{o}(\rho)\leq\mathbb{S}_{o}(\overline{\rho})$, which indicates that $\mathbb{S}_{o}(\overline{\rho})=\mathbb{S}_{o}(\rho)$. So the quantity $\mathbb{S}_{o}$ does not change under $LU$ operations.\qed\\

\end{document}